\documentclass{article}

\usepackage[preprint]{spconf}

\usepackage{amsmath,graphicx}
\usepackage{tabularx,tabulary,booktabs}
\usepackage{float}

\usepackage{enumitem}
\setlist{nosep, leftmargin=14pt}

\usepackage{mwe} 

\copyrightnotice{\copyright\ IEEE 2021}
                \toappear{To appear in {\it Proc.\ ISBI 2021, April 13-16, 2021, Nice, France}}

\newcommand\blfootnote[1]{%
  \begingroup
  \renewcommand\thefootnote{}\footnote{#1}%
  \addtocounter{footnote}{-1}%
  \endgroup
}

\title{Learning Multi-Modal Volumetric Prostate Registration with Weak Inter-Subject Spatial Correspondence}
\name{
\begin{tabular}{c}Oleksii Bashkanov$^{\star}$ \qquad Anneke Meyer$^{\star}$ \qquad  Daniel Schindele$^{\dagger}$ \qquad Martin Schostak$^{\dagger}$ \\ \qquad  Klaus–Dietz Tönnies$^{\star}$ \qquad  Christian Hansen$^{\star}$ \qquad  Marko Rak$^{\star}$
\end{tabular}}

\address{$^{\star}$ 
Faculty of Computer Science \& Research Campus STIMULATE, University of Magdeburg, Germany \\
$^{\dagger}$ Clinic of Urology and Pediatric Urology, University Hospital Magdeburg, Germany}

\begin{document}
\ninept 

\maketitle
\begin{abstract}
Recent studies demonstrated the eligibility of convolutional neural networks (CNNs) for solving the image registration problem. CNNs enable faster transformation estimation and greater generalization capability needed for better support during medical interventions. Conventional fully-supervised training requires a lot of high-quality ground truth data such as voxel-to-voxel transformations, which typically are attained in a too tedious and error-prone manner. In our work, we use weakly-supervised learning, which optimizes the model indirectly only via segmentation masks that are a more accessible ground truth than the deformation fields. Concerning the weak supervision, we investigate two segmentation similarity measures: multiscale Dice similarity coefficient (mDSC) and the similarity between segmentation-derived signed distance maps (SDMs). We show that the combination of mDSC and SDM similarity measures results in a more accurate and natural transformation pattern together with a stronger gradient coverage. Furthermore, we introduce an auxiliary input to the neural network for the prior information about the prostate location in the MR sequence, which mostly is available preoperatively.
This approach significantly outperforms the standard two-input models. 
With weakly labelled MR-TRUS prostate data, we showed registration quality comparable to the state-of-the-art deep learning-based method.
\end{abstract}

\blfootnote{
\copyright 2021 IEEE. Personal use of this material is permitted. Permission from IEEE must be obtained for all other uses, in any current or future media, including reprinting/republishing this material for advertising or promotional purposes, creating new collective works, for resale or redistribution to servers or lists, or reuse of any copyrighted component of this work in other works.}

\section{Introduction}
\label{sec:intro}

Prostate cancer is the second most prevalent cancer diagnosis among older men. A recent review \cite{rawla2019epidemiology} reports that 1,276,106 new cases and 358,989 deaths, were registered in 2018 worldwide. Early-stage diagnosis is one of the essential countermeasures, easing treatment of prostate cancer, and lowering mortality risks. Medical image registration plays a significant role in this context, supporting physicians during their image-guided biopsies and therapies. Transrectal ultrasound (TRUS) biopsies are among the most relevant techniques for early diagnosis of prostate cancer. However, TRUS alone does not provide the necessary soft-tissue contrast. Therefore, biopsies are usually preceded by magnetic resonance (MR) imaging to identify suspicious regions beforehand. During the actual biopsy procedure, the information from MR and TRUS needs to be registered in real-time to steer the tissue sampling.

The main challenge of the underlying registration problem is the massive difference in the appearance of the prostate between MR and TRUS sequences, causing many classical iterative intensity-based methods to fail. The alternative would be to use segmentation-driven registration methods. While MR image segmentation is done preoperatively, the TRUS sequence requires a manual intra-procedural segmentation which prolongs the biopsy. A recent practical survey \cite{andrade2018practical} revealed that CNNs are the most promising way to address this challenge, enabling real-time MR-TRUS prostate registration without any additional manual guidance or effort.

Early research focused on segmentation-driven registration. For instance, \cite{fedorov20123d} presented two techniques: B-spline registration of segmentation-derived signed distance maps and a biomechanically constrained surface registration. In both cases, an intra-procedural prostate segmentation is required. Since many intensity-based similarity measures are prone to fail in multi-modal MR-TRUS settings, \cite{haskins2019learning} proposed to learn image similarity through CNNs, outperforming mutual information on MR-TRUS pairs. However, this approach has its drawbacks. The learned similarity measure needs to be enveloped by an iterative registration which contradicts the real-time requirement. Moreover, the training of similarity measures requires that voxel-correspondence is available as ground truth, which is challenging to collect. This is especially true for TRUS because prostate deformations are involved due to the transrectal ultrasound transducer.

To address the ground truth collection problem, \cite{krebs2017robust} proposed an agent-based generation of synthetic ground truth for deformable registration, which can be jointly used with real ground truth to train prostate registration in MR-based setting. 
However, the strategies already exist to avoid dense voxel-correspondence altogether.
To this end, \cite{hu2018label, hu2018weakly} proposed to use a weakly-supervised technique, combining sparse correspondence from anatomical landmarks and prostate gland segmentations into a weaker proxy measure.
This neat ideas reduce the effort involved in the creation of the ground truth data, but still requires tedious segmentation of the corresponding prostate anatomical landmarks on both sequences.

In this work, we go one step further, proposing to only use the available prostate gland segmentations as a weak proxy measure 
between two modalities. 
This approach greatly reduces the effort spent in the creation of the ground truth data compared to the voxel-to-voxel correspondence or the patient-specific landmark pairs.
Apart from that, the multitude of research behind automatic prostate segmentation indicates a great potential for automating this task \cite{meyer2018automatic, wang2019deep, aldoj2020automatic}. This can enable automatic annotation of  massive datasets, where human input would be merely impractical on such a scale.

We compare three approaches of utilizing segmentation similarity measures to optimize registration models based on  mDSC, the similarity between segmentation-derived SDMs and on the combination of them.
Moreover, we show that the prior information about the prostate location on the preoperative MR sequences, for which segmentations are mostly available beforehand, can significantly improve the overall registration performance.

\begin{figure}[b]
      \centering
      \includegraphics[trim={3.7cm 0.0cm 0.0cm 0.0cm}, width=1\linewidth]{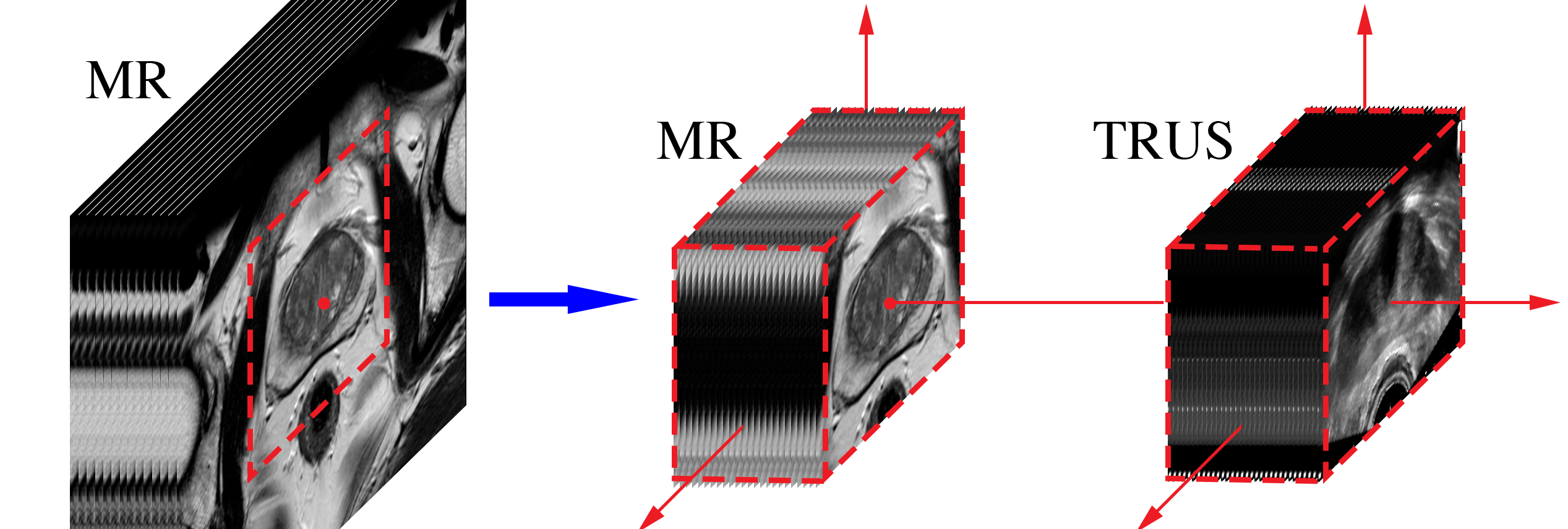}
      \caption{\small Illustration of the coarse alignment via region of interest definition on MR sequence based on center of mass according to MRI segmentation and the target image size derived from TRUS volume.}
      \label{fig:axis}
\end{figure}

\section{Methods}
Our pipeline consists of two stages: an initial coarse alignment of the regions of interests and a subsequent deformable fine registration. In this section, we provide the details for both stages.

\subsection{Coarse pre-alignment}
In the first stage, we seek to guarantee a coarse alignment of both images. To this end, we noticed that in both images, the prostate is oriented according to the axis of the coordinate system \mbox{(see Fig. \ref{fig:axis})}, meaning that there is little to no rotation involved. Therefore, we could use the axis as a reference frame for coarse initialization. Regarding translation, we found that for TRUS, the prostate is approximately centrally positioned on the image due to acquisition. This needs not to be always true for the MR images. To compensate for the difference, we used the available MR-based prostate segmentation to align its center of mass with the TRUS centre. After the alignment, we resampled the MR according to the TRUS image and cropped both images to guarantee a fixed input size for our CNN. Typically, the MR segmentation is done by a radiologist before the biopsy procedure; however, this step can be automated easily, for example, using U-Net-based techniques \cite{meyer2018automatic} in this stage.

\subsection{Fine alignment} In the second stage, a deformation field is predicted to establish the spatial correspondence between voxels of the MR (moving) and TRUS (fixed) image using a one-step CNN. We trained our network in weakly-supervised fashion. Specifically, the training was driven by the overlap between the labels of the moving and fixed image.

Standard binary overlap measures like the Dice similarity coefficient (DSC), widely used for segmentation problems, can fail for registration problems. For instance, in case of a complete label mismatch, any network will struggle to adapt its parameters due to zero gradients. To overcome this problem, we used the method presented in \cite{hu2018weakly} which uses a multiscale DSC (mDSC), that averages over a battery of subsequently Gaussian smoothed binary masks. Specifically, we used
\begin{equation}
    \mathcal{L_{DSC}}(p,g) =  \dfrac{1}{Z}\sum\limits_{z \in \sigma} \dfrac{2 \sum_{i}^{N} p_{z,i} \cdot g_{z,i}}{\sum_{i}^{N} p_{z,i} + \sum_{i}^{N} g_{z,i}},
\end{equation}
\noindent where $Z$ is the number of smoothness levels, and $p$ and $g$ being the registered moving and fixed binary segmentation accordingly. 

We utilized Gaussians filters of sizes $\sigma \in \{0, 1, 2, 4, 8\}$. The special case $\sigma=0$ indicates that no smoothing is applied, which suggests the standard DSC definition. The idea is that larger Gaussians will contribute more to the global convergence of the deformation, while smaller Gaussians will carve out the fine detail. A useful by-product of this mDSC definition is the inherent smoothness of the deformation.

\begin{figure}
      \centering
      \includegraphics[trim={3.5cm 8.9cm 2.5cm 0.7cm}, width=.96\linewidth]{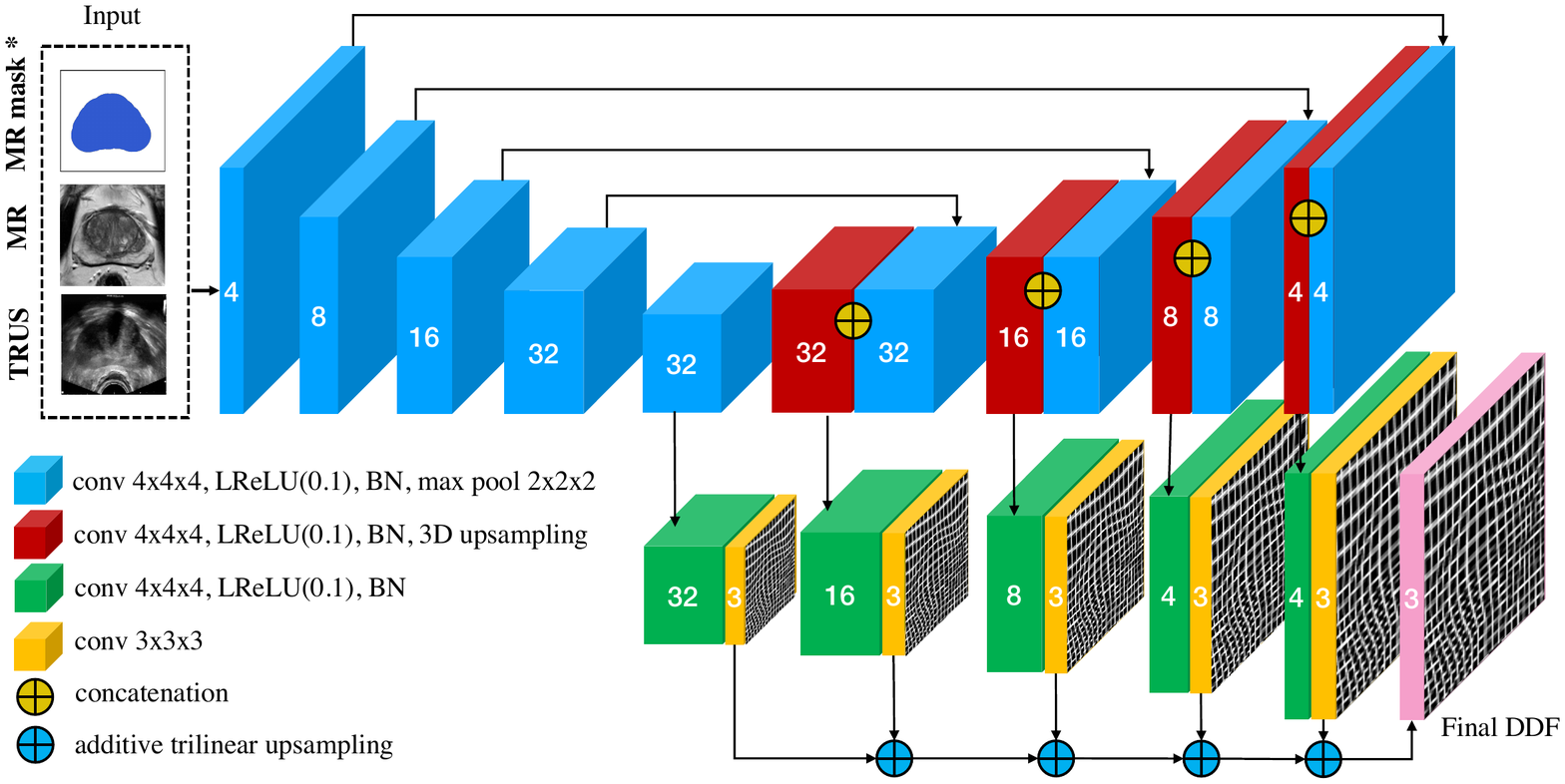}
      \caption{\small Network architectures with a cumulative summation of the transformation from the coarse- up to the full-scale level. \text{*}The third input for MR segmentation is optional.}
      \label{fig:unet}
\end{figure}
The add-on to regular DSC mentioned above is unable to cover the whole deformable region with strong gradients entirely. To overcome this problem, we propose to use segmentation-derived euclidean signed distance maps (SDMs) as a driving force to increase the overlap between $p$ and $g$. This strategy has already been shown to be favourable in the context of iterative prostate registration \cite{fedorov20123d}. 
In our case, we sought to minimize the mean squared logarithmic error (MSLE) between pre-computed deformed $\hat{p}$ and fixed $\hat{g}$ SDMs. MSLE operates on all positive sides of the SDMs and thus propagates the gradients to the area outside the prostate. Presumably, the resultant deformation field should imitate how the prostate is deformed during the procedure --- with external forces applied to it. Another useful property is that the regions closer to the prostate boundary will be penalized stronger.
\begin{equation}
     \mathcal{L_{SDM}}(\hat{p}, \hat{g}) = \frac{1}{N}\sum_{i=1}^{N}(\log(\hat{p_{i}} + 1)-\log(\hat{g_{i}} + 1))^2
\end{equation}
To ensure the smoothness of deformations, we additionally applied bending energy \cite{rueckert1999nonrigid} as a second-order smoothness penalty $\mathcal{T}$ for volumetric dense deformation field (DDF) $\boldsymbol{u}$, where only non-linear deformations are penalized while allowing global affine transformations:
\begin{equation}
\resizebox{1.0\hsize}{!}{$
    \mathcal{T}(\boldsymbol{u}) = \int_{0}^{x} \int_{0}^{y} \int_{0}^{z} \left[ \left(\frac{\partial^{2} \boldsymbol{u}}{\partial x^2}\right)^2 + \left(\frac{\partial^{2}\boldsymbol{u}}{\partial y^2}\right)^2 + \left(\frac{\partial^{2}\boldsymbol{u}}{\partial z^2}\right)^2  \right]\partial x \: \partial y \: \partial z.$
}
\end{equation}

\begin{figure*}[h]
      \centering
      \includegraphics[width=.99\textwidth]{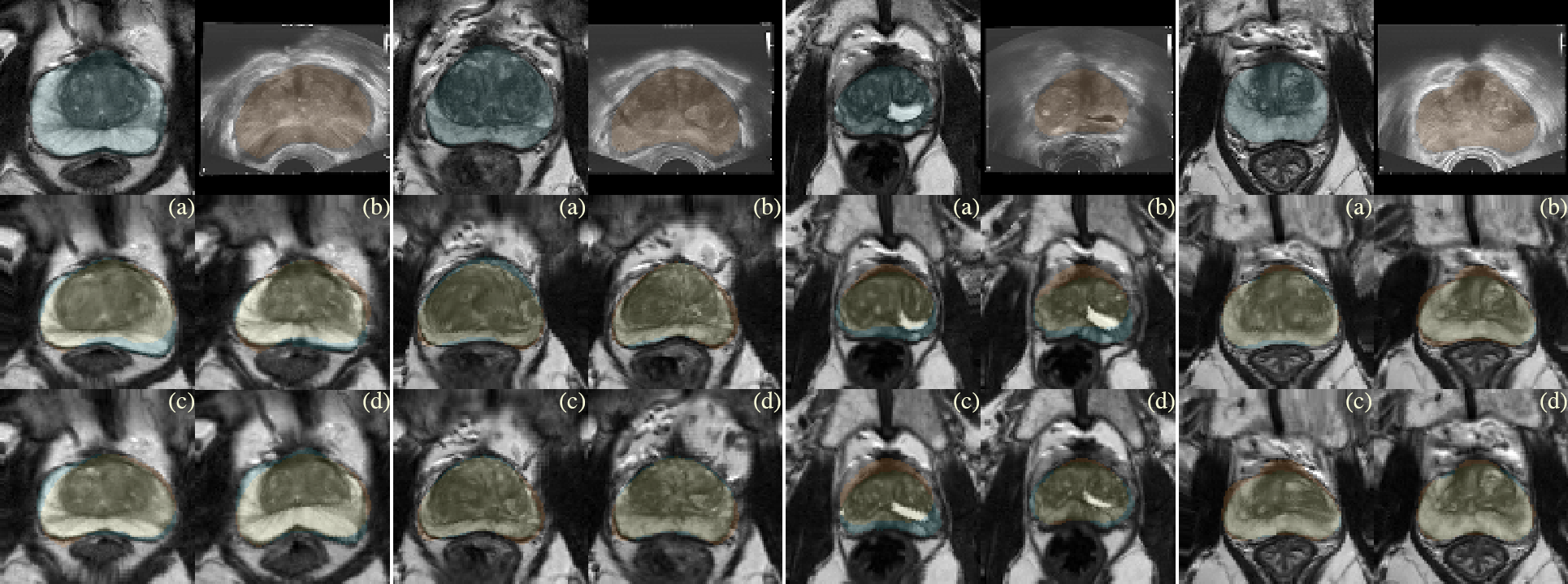}
      \caption{\small Qualitative comparison of the proposed approaches: (a) multiscale Dice similarity coefficient (mDSC), (b) signed distance maps (SDM), (c) mixed strategy (MIX), and (d) its variation with mask (MIX+). Each 3$\times$2 block depicts individual case. Top row contains pre-aligned MR-TRUS pairs. The slices were taken in axial plane with MR segmentation (cyan), TRUS segmentation (orange), and segmentation overlap (yellow).}
      \label{fig:mrtrusimgs}
\end{figure*}

\subsection{Implementation} We utilized a 3D U-Net-like architecture (Fig.~\ref{fig:unet}), actively used in the medical imaging analysis \cite{cciccek20163d}. The first part of this architecture comprises downsampling levels, which make the resolution smaller while expanding the feature space. The second part performs upsampling from high-dimensional feature space until it reaches the level with the same input resolution. Between these two parts, there are skip connections at each corresponding level, which provide supportive high-resolution information for the upsampling part.

Similarly to \cite{hu2018weakly}, we exploited the idea of the additive upsampling in multi-resolution way. This allows the network to decouple the global and fine transformations and learn them separately. A DDF is produced at each level by the 3D convolution block branched from the layer preceding upsampling operation (Fig.~\ref{fig:unet}). To get to the final DDF, we resample the coarser outputs to the next higher resolution level and add them up. We do this progressively until we reach the full resolution of the displacement.
Put differently, our network combines DDFs in a pyramidal fashion at $1, 1/2, 1/4, 1/8$ and $1/16$ resolution scaling levels.

Our network architecture comprises 3D convolutional layers with the kernel size of $4 \times 4 \times 4$, the stride of $1 \times  1 \times 1$, and the same padding for the down- and upsampling part. At the first level, the convolution layer outputs four filters, followed by Leaky ReLU activation with a negative slope of 0.1 with the subsequent batch normalization layer. Later, this layer is downsampled by 3D MaxPooling operation with a stride of $2 \times 2 \times 2$. With every next level moving down, the number of filters in the feature space is doubled (see Fig.~\ref{fig:unet}). The residual convolution layers from the upsampling part share the kernel size of $3 \times 3 \times 3$.


We have trained our models with a total batch size of 5. The Adam optimization algorithm was used with an initial learning rate of $2e-4$. All configurations were trained for 300 epochs at most (33,000 mini-batch iterations), while best models were selected on the best validation DSC. 3D affine image augmentation, together with a trilinear resampling layer, and differentiable deformation layer was borrowed from \cite{hu2018weakly}, which is an adapted version of open-source methods from NiftyNet \cite{gibson2018niftynet}. 

\subsubsection{Data} Our dataset contained 155 segmented TRUS and T2-weighted MR image pairs. For each patient, these MR records were received in a transversal scanning position from a Philips Achieva 3T system with a volume size of $320\times320\times30$ voxels and resolution of $0.5\times0.5\times2.75$ mm. For biopsies, the BK 3000 ultrasound system, together with an endorectal biplane transducer, was used to acquire image sequences of size $140\times115\times120$ with $0.6\times0.6\times0.6$ mm spacing. While the prostate glands from MRI scans were segmented semi-automatically by the radiologist with the commercial software Philips DynaCAD beforehand, the TRUS segmentation masks were acquired during biopsy intervention in a semi-supervised manner using the commercial software Philips UroNav. Since not all TRUS segmentations were of the best quality due to intra-procedural acquisition, 21 held-out test cases were thoroughly re-segmented on the TRUS images for reliable evaluation. Moreover, 73 new anatomical landmark pairs in the form of patient-specific calcification and cysts were annotated and verified by an experienced urologist on this data for a detailed evaluation of the methods.

\subsubsection{Pre-processing} Both MR and TRUS images were resampled to the voxel size of $0.88 \times 0.88 \times 0.88$ mm to place them into the same world coordinate system. For the resampling procedure on the image data, we utilized B-Spline interpolation, whereas, for segmentation masks, linear interpolation followed by binarization operation was used. Later, all images were cropped to the volume size of $96\times96\times80$. Subsequently, both TRUS and MR image intensities up to the 99th percentile were normalized to the interval of $[0, 1]$. During the training, we separately augmented the input data on the fly by resampling the TRUS and MR images with a small randomly generated affine transformation without flipping.

\subsection{Experiments}
We performed several experiments with five-fold cross-validation scheme showing that weakly-supervised learning strategy can achieve state-of-the-art performance in automatic transformation estimation task with a considerably reduced time and effort spent in creation of the ground truth data. 

First, we aimed to investigate how the mDSC- and SDM-driven registration methods perform with registration task individually. Then, we examined the idea of combining these two approaches and how it impacts the deformation behaviour.
Lastly, we compared the registration accuracy between the models with and without the third auxiliary input for MR segmentation. We hypothesize that this prior information provides substantial support in identifying the prostate location on moving images. 


\subsubsection{Loss function}
In our experiments, the loss function combines two label similarity metrics with the smoothness penalty:
\begin{equation}
    \mathcal{L} = \alpha \mathcal{L_{DSC}}  + \beta \mathcal{L_{SDM}} + \gamma\mathcal{T},
\end{equation} where $\mathcal{L_{DSC}}$ and $\mathcal{L_{SDM}}$ denote the mDSC score and the similarity error between SDMs, respectively. Bending energy $\mathcal{T}$ is weighted by \mbox{$\gamma$ term,} with \mbox{$\gamma=1-\alpha-\beta$}. To compare different strategies we trained models independently with empirically chosen hyper-parameters $\alpha=0.3, \beta=0$ for mDSC and $\alpha=0, \beta=0.8$ for SDM. The final mixed strategy (MIX) combines multiscale DSC and similarity of SDMs with the following parameterization: $\alpha=0.05$ and $\beta=0.45$, since it yielded the best results.

\section{Results and discussion}

\addtolength{\tabcolsep}{-1pt}    

\begin{table}[b]
\caption{\small Quantitative comparison of the registration methods in terms of Dice similarity coefficient (DSC), Target registration error (TRE) and the mean gradient of the Jacobian determinant $\times10^{-2}$. The results are in the $mean(median) \pm SD$ format. Symbol $\uparrow$ indicates that a lower score is better, while $\downarrow$ vice versa.}
\label{tab:quantitative}

\smallskip
\centering
\begin{tabularx}{1.\columnwidth}{X|ccc}

\toprule[\heavyrulewidth]\toprule[\heavyrulewidth]

\bf Method   &  DSC Whole $\uparrow$     &  TRE, mm $\downarrow$ & $\nabla|J|\times10^{-2}$ $\downarrow$  \\ 
\midrule
\textit{Coarse}   & 74.3 (76.0) $\pm$ 10.6 & 7.0 (7.4) $\pm$ 2.8 &  - \\

mDSC     & 86.5 (87.2) $\pm$ 4.0 & 6.1 (5.9) $\pm$ 2.1 &     30 (30) $\pm$ 5   \\
SDM      & 85.7 (87.1) $\pm$ 5.2 & 5.3 (4.9) $\pm$ 2.3 &      65 (64) $\pm$ 9   \\
MIX      & 85.1 (87.0) $\pm$ 5.2 & 5.4 (4.8) $\pm$ 2.2 &      \textbf{28 (27) $\pm$ 4}    \\
MIX+     & \textbf{88.3 (89.6) $\pm$ 3.8} & \textbf{4.7 (4.3) $\pm$ 1.9} &      36 (36) $\pm$ 5    \\

\bottomrule[\heavyrulewidth]


\end{tabularx}
\end{table}

For each experiment, we measured the registration accuracy with the DSC score on the whole prostate gland and its three anatomical regions (base, mid, apex) together with patient-based TRE by calculating the mean distance between the center of mass of the corresponding anatomical landmarks. The coarse pre-alignment on the test data produced a median DSC score of 0.76 and median TRE of 7.4 mm, which is sufficient for initialization stage.

\begin{figure}[t]
      \centering
      \includegraphics[trim={1.4cm 0.9cm 0.15cm 0.5cm},width=.999\linewidth]{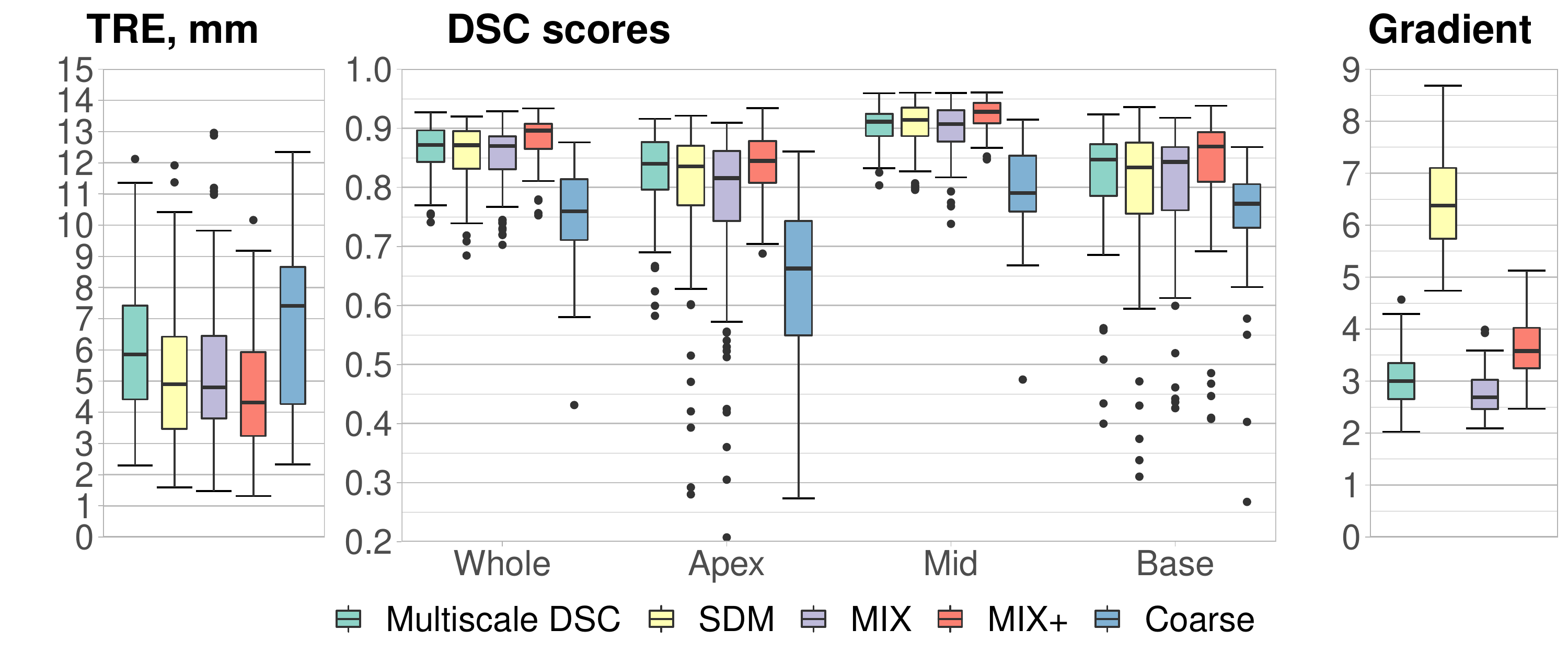}
      \caption{\small Registration performance comparison of multiscale Dice similarity coefficient (mDSC), signed distance maps (SDM) and mixed strategy (MIX) together with its variation with mask (MIX+) in terms of Dice similarity coefficient (DSC), Target registration error (TRE) and the mean gradient of the Jacobian determinant $\times10^{-2}$. Coarse alignment indicates starting conditions.}
      \label{fig:boxplots}
\end{figure}

Visual inspection revealed that mDSC-based models tend to stretch the inner organ structures to its boundaries, whereas the SDM-based models demonstrated a more uniform and natural deformation pattern (Fig. \ref{fig:mrtrusimgs}), also resulting in a better TRE score. However, the SDM model fails to produce smooth deformation according to the second-order smoothness measure $\nabla|J|$ (see Fig.~\ref{fig:boxplots}). Since mDSC approach entails the regularization effect on the DDF, it is reasonable to try to combine it with SDM. The performance score of such approach indicates that MIX learning yields a much smoother transformation than SDMs while preserving the high registration accuracy.

Considering the auxiliary input of the network (MIX+), we found that the additional information in the form of the MR segmentation mask helped to achieve significantly better results for DSC and TRE than the standard two-images input (p-value of $< 0.001$ for both metrics in a paired Wilcoxon signed-rank test).
These findings suggest that with such prior information, the network can more easily figure out \textit{where} to move, without looking for \textit{what} to move.

In general, our MIX+ model with auxiliary input reached state-of-the-art performance with the median TRE of 4.3 mm and DSC of 0.89, which is similar to the results demonstrated by Hu et al. \cite{hu2018weakly} with TRE of 3.6 mm and DSC of 0.88. In contrast to our method, they utilize plenty of anatomical landmarks during training, whereas our training relied purely on the fixed and moving segmentation masks. It is also worth noting that up to this point there is no similar dataset publicly available, that would contain such MR-TRUS pairs per patient and enable more transparent comparison.

In future work, it is reasonable to investigate the impact of the deep-learning generated segmentation masks on the registration performance. Thus, for instance, the coarse pre-alignment can be improved by having two segmentations during inference stage instead of one MR mask as it is now. Consequently, a better starting position can lead to more accurate fine registration. This will also enable the comparison of the segmentation-based iterative registration method on automatically created segmentation with the instant DDF prediction methods. 
\section{Conclusion}
This work demonstrated that it is feasible to learn an effective registration model based solely on the segmentation of the prostate glands. The combination of two segmentation similarity measures (mDSC and SDM) has proven to be the best option in terms of accuracy and smoothness of the transformation. We demonstrated that models with the given prior information, such as MR prostate segmentation, significantly outperform the two-input models.

Our approach enables efficient registration learning without the use of anatomical landmarks and requires far less ground truth data than the fully-supervised methods. Future work should also focus on the potential integration of the deep-learning-based intensity-based similarity metric \cite{haskins2019learning} in the current setup to eliminate bias imposed during the segmentation process and explore the effects of DDF ensembling.

\paragraph*{Acknowledgments}
\label{sec:acknowledgments}
This work has been supported by the federal state of Saxony-Anhalt, Germany within the framework of the postgraduates funding. We would like to thank all the colleagues and radiologists, who have contributed their effort to this study.

\paragraph*{Compliance with Ethical Standards}

The retrospective analysis in this research study relies on fully anonymized treatment planning data. This work does not involve any studies with human participants or animals performed by any of the authors and is in line with the Declaration of Helsinki.

\paragraph*{Conflict of Interest}
The authors have no conflicts of interest to declare.
\bibliographystyle{IEEEbib}
\bibliography{strings}

\end{document}